\documentclass[universe,a4paper,essay,accept,pdftex,moreauthors]{Definitions/mdpi} 

\newcommand{\be}{\begin{equation}}
\newcommand{\ee}{\end{equation}}
\newcommand{\ba}{\begin{eqnarray}}
\newcommand{\ea}{\end{eqnarray}}

\newcommand{\R}{{\mathbb R}}
\newcommand{\Z}{{\mathbb{C}}}
\newcommand{\Ha}{{\mathbb H}}
\newcommand{\Oc}{{\mathbb O}}
\newcommand{\algebras}{\R\otimes\Z\otimes\Ha\otimes\Oc}
\newcommand{\eh}{\text{e}}
\newcommand{\eo}{\text{f}}
\newcommand{\ii}{\text{i}}

\newcommand{\norm}[1]{\| #1\|}

\def\d{\mathrm{d}}

\newcommand{\nuM}{\hat{\nu}}
\newcommand{\cM}{\hat{C}}
\newcommand{\pM}{\hat{N}}
\newcommand{\mM}{\hat{M}}

\newcommand{\ssl}{\mathfrak{sl}}

\firstpage{1} 
\pubvolume{8}
\issuenum{8}
\articlenumber{0407}
\pubyear{2022}
\copyrightyear{2022}
\datereceived{19 May 2022} 
\dateaccepted{28 July 2022} 
\datepublished{3 August 2022} 
\hreflink{} 

\pdfoutput=1

\Title{Listening to Celestial Algebras}

\Author{{Jose} 
 Beltr\'an Jim\'enez $^{1}$\orcidA{} 
 and Tomi S. Koivisto $^{2,3}$\orcidB{}
 }

\address{%
$^{1}$ \quad Departamento de F\'isica Fundamental and IUFFyM, Universidad de Salamanca, {37008} 
 Salamanca, Spain;\\
$^{2}$ \quad Laboratory of Theoretical Physics, Institute of Physics, University of Tartu, W. Ostwaldi 1, 50411 Tartu, Estonia.\\
$^{3}$ \quad National Institute of Chemical Physics and Biophysics, Rävala pst. 10, 10143 Tallinn, Estonia.}

\corres{\href{mailto:jose.beltran@usal.es}{jose.beltran@usal.es}; \href{mailto:tomi.koivisto@ut.ee}{tomi.koivisto@ut.ee}}

\abstract{\textls[-25]{In this essay, we immerse into the framework of normed division algebras as a suitable arena to accommodate the standard model of elementary particles, and we explore some applications to cosmology. Remarkably, they permit interesting non-trivial realisations of the cosmological principle with an interplay between the symmetry groups of the quaternions and octonions. We also argue how these realisations give rise to potentially observational signatures in gravitational waves astronomy.}}


\begin{document}

\section{Introduction}
One of the wonders of the natural world resides in its compliance to being described in a mathematical language. In~this language, numbers form the channel to communicate with Nature, and one could even assert that science (and physics in particular) is about obtaining theoretical predictions expressed in terms of numbers that can eventually be compared to observational measurements. Natural numbers allow us to count things, and that is why they were the first ones to make an appearance as a practical construct. Soon, negative natural and rational numbers claimed their position for practical purposes. Beyond~this point, it would seem that nothing more was necessary for the primordial mathematical needs of the first societies. However, when geometry had to be included in the mathematical machinery, new numbers such as $\sqrt{2}$ and $\pi$ also called their role to measure lengths and surfaces. Subsequent needs and mathematical consistency gave the leading role in this play to the real numbers $\R$ and, not coincidentally, they were the first field to be discovered. They are a continuum completion of the more intuitive rational numbers and form an algebra that possesses all the nice properties that one could ask for, namely: it is ordered, commutative and associative. Moreover, one can define a norm and a well-defined division, thus making it a normed division algebra.  Nevertheless, the reals fail in one important aspect: polynomial equations with real coefficients do not necessarily admit real solutions. This innocent observation prompted the introduction of the imaginary unit $\ii$ and led to the discovery of the complex numbers $\Z$ that extend $\R$ into a closed field so that all polynomial equations now find solutions within the same field, which is a~result known as the fundamental theorem of algebra proven by Carl Friedrich Gauss. The~complex numbers share the commutativity and associativity properties of the reals as well as maintaining the status of a normed division algebra. They lose the order of the reals, but~this is a hitch we are willing to accept in view of all the additional gifts brought about by them. Despite appearing as a mathematical construct, it is undeniable that Nature embraces the complex numbers, and they play a paramount role in fundamental aspects as quantum mechanics,\footnote{We can allude to the Aharonov--Bohm effect as a {\it real} manifestation of complex numbers in Nature.} but also in very practical situations such as the description of electric~circuits.

After discovering the reals and the complex numbers and their suitability to describe Nature, a~pertinent question would be: are there more types of numbers that could help us in our understanding of the natural world? If so, is there an infinite class of {\it{ numbers}} that will appear in our ever-increasingly fundamental comprehension of the natural laws? The answer to these questions is intriguingly precise: there are exactly four distinctive types of numbers (conforming normed division algebras). However,~then, does this mean that they suffice to have a complete fundamental description of Nature? In the following, we will occupy ourselves with the first questions and defer this more ambitious last question to another~occasion.

\section{The neglected but valuable~quaternions}

The search for numbers beyond the complex realm commenced from a pure mathematical curiosity in an endeavour undertaken by William Rowan Hamilton, who discovered the quaternions $\Ha$ as a four-dimensional extension of the complex numbers: a~discovery that was literally carved in stone for the posterity in the Brougham Bridge of Dublin. These numbers still form a normed division algebra, but~they leave the commutativity of $\Z$ behind. The~quaternions extend the complex numbers with two additional imaginary units so an arbitrary $q\in\Ha$ can be expressed as
\be
q=q_0+q_1 \eh_1+q_2 \eh_2+q_3 \eh_3\,,
\ee
where $\eh_i$ belong to a set that generates, together with the unit element $1$, the~quaternionic algebra $\Ha=\text{Span}\{1,\eh_1,\eh_2,\eh_3\}$ via the following product rules:
\be
\eh_i\eh_j=-\delta_{ij}+\epsilon_{ijk}\eh_k,
\label{eq:productquat}
\ee
that give rise to the commutation relations
\be
[\eh_i,\eh_j]=2\epsilon_{ijk}\eh_k.
\ee

These commutation relations already hint at the suitability of quaternions to describe three-dimensional rotations in terms of multiplication of pure imaginary quaternions with $q_0=0$. We can also introduce the conjugation that consists in changing the sign of the imaginary units: $\bar{q}\equiv q_0q_0-q_1 \eh_1-q_2 \eh_2-q_3 \eh_3$, so a quaternion is said to be purely imaginary if it satisfies $\bar{q}=-q$. The~norm of a quaternion is then defined as $\norm{q}^2=q\bar{q}=q_0^2+q_1^2+q_2^2+q_3^2$ and its inverse $q^{-1}=\bar{q}/\norm{q}$.

At the time, the~quaternions were a very fashionable subject, and physics made extensive use of them (e.g., Maxwell equations were written with quaternions). When Gibbs noticed that quaternions could be expressed in the three-dimensional vector space endowed with a dot and a cross product, they became a proper contender. Quaternions eventually lost the battle and vectors arouse as the standard framework for physics, displacing quaternions to a marginal place. They however have always lurked in different corners of physics, finding their way to provide insightful applications. One can easily understand why they are an appealing groundwork for physics after noticing that they admit a representation in $\text{M}(2,\Z)$ via
\be
q\mapsto
A(q)=\begin{pmatrix}
q_0+\ii q_1 & q_2+\ii~q_3 \\
-q_2+\ii q_3 & q_0-\ii q_1
\end{pmatrix}\,,
\label{eq:Ahom}
\ee
where multiplication in $\Ha$ simply becomes matrix product with $A(q_1q_2)=A(q_1) A(q_2)$, i.e.,~the above gives an homomorphism between quaternions and $2\times2$ complex matrices. One can observe that imaginary quaternions can be expressed in terms of Pauli matrices, thus corroborating their relation to rotations. Furthermore, the~homomorphism also shows that $\norm{q}^2=\det A(q)$, which further unveils the isomorphism between unit quaternions and $SU(2)$. It is then direct to uncover that rotations can be realised as unitary quaternions $r$ (which satisfy $\bar{r}=r^{-1}$) acting on pure imaginary quaternions $x$ (that satisfy $\bar{x}=-x$ by definition) via $x\mapsto r\,x\,\bar{r}$.

It is certainly very appealing that pure imaginary quaternions provide a realisation of the real three-vector space where rotations is simply realised by quaternionic multiplication as well as their intimate relation to $SU(2)$. In~view of these properties and noticing that quaternions are actually a four-dimensional algebra, which coincides with the spacetime dimension, one cannot help but wonder if the spacetime Lorentz symmetry can find a dwell within the quaternionic algebra. The~fascinating answer is that it indeed does! Perhaps even more fascinating is that Lorentz symmetry rightfully finds its place not in $\Ha$ but in $\Z\otimes\Ha$. After~all, why should we leave our old friend $\Z$ out of the function? 

The application of complex quaternions to special relativity did not take long since its inception by Einstein in 1905. In~two independent works by A. Conway in 1911~\cite{Conway1911} and L. Silberstein a year later~\cite{SilbersteinLXXVIQF}, special relativity and Lorentz transformations were presented to quaternions, and it was explored in subsequent years by different authors (see e.g.,~\cite{Synge:1972zz} and references therein). We can show how to realise spacetime Lorentz symmetry by identifying the spacetime position with the quaternion 
\be
x=-\ii x^0+x^i\eh_i,
\ee
that is antihermitian $x+x^{\dagger}=0$, where $x^\dagger$ is the complex and quaternion conjugate of $x$. It is straightforward to check that this condition is preserved under the quaternion transformation $x\mapsto \hat{q}\,x\,\hat{q}^{\dagger}$ with $\hat{q}$ a unit complex quaternion. Moreover, the~norm \linebreak $\norm{x}^2=-(x^0)^2+\vec{x}^2$, that gives the spacetime interval, is also invariant. This permits realising Lorentz transformations of the spacetime coordinates as a multiplication by unit complex quaternions in the space of antihermitian complex quaternions (see e.g.,~\cite{Rao1983} for an explicit construction). Restricting to real unit quaternions, we recover the spatial rotations, so boosts are naturally associated to the antihermitian (pure quaternionic imaginary) part of $\hat{q}$. 

Very much like pure imaginary quaternions over the reals realise rotations in the Euclidean space, pure imaginary quaternions with complex coefficients generate the Lie algebra $\ssl(2,\Z)$, which precisely corresponds to the double cover of the Lorentz group. This can be easily seen from \eqref{eq:Ahom} with $q_0=0$. If~we introduce the basis
\be
\left\{\hat{\eh}_1=\ii \eh_1,\;\hat{\eh}_2=\frac{1}{\sqrt{2}}(\eh_2+\ii\eh_3),\;\hat{\eh}_3=-\frac{1}{\sqrt{2}}(\eh_2-\ii\eh_3)\right\}\,,
\ee
it is immediate to obtain
\be
q=z_1\hat{\eh}_1+z_2\hat{\eh}_2+z_3\hat{\eh}_3\quad\mapsto
\quad A(q)=\begin{pmatrix}
z_1 & z_2 \\
z_3 & -z_1
\end{pmatrix},\quad\quad z_1,z_2,z_3\in\Z,
\label{eq:Ahomtosl}
\ee
that establishes the realisation of $\ssl(2,\Z)$ with unit quaternions. This should be sufficiently convincing and stimulating to embrace the algebra $\Z\otimes\Ha$ as a promising framework to describe the Lorentz group. Of~course, a number of authors have engaged this venture with interesting results. A~related subject that deserves attention is that once we have a proper characterisation of the Lorentz group within $\Z\otimes\Ha$, the~possibility of describing gravity as a localisation of this algebra emerges as~well as new avenues to exploring theories of gravity. This is a speculation we should not dismiss and should receive a more meticulous~scrutiny.

\section{The surprisingly cooperative character of the untamed~octonions}

Having met the quaternions and its interesting applications, our curiosity eagerly craves the exploration of more algebras and their suitability to describe physics. The~task of obtaining higher-dimensional extensions of the quaternions was tackled by John T. Graves, who showed the existence of an eight-dimensional algebra that he called {\it octaves} but~are now known as {\it octonions}. This algebra is generated by
\be
\Oc={\text{Span}}\{1,\eo_1,\eo_2,\eo_3,\eo_4,\eo_5,\eo_6,\eo_7\}
\ee
that satisfy the following~rules:
\begin{itemize}
    \item $\eo_a^2=-1$.
    \item Anticommutativity: $\{\eo_a,\eo_b\}=0$ for $a\neq b$.
    \item Cycling identity: If $\eo_a\eo_b=\eo_c$, then $\eo_{a+1}\eo_{b+1}=\eo_{c+1}$ mod 7.
    \item Index doubling identity: If $\eo_a\eo_b=\eo_c$, then $\eo_{2a}\eo_{2b}=\eo_{2c}$ mod 7.
\end{itemize}

The octonion multiplication table is not particularly illuminating. For~a detailed and excellent review of the properties and some applications of the octonions, we refer to~\cite{Baez:2001dm}. In~analogy with the quaternions, we can write the octonions multiplication as
\be
\eo_a\eo_b=-\delta_{ab}+f_{abc}\eo_c
\ee
where $f_{abc}$ represents the structure constants of the octonions algebra, which are completely antisymmetric. A~common representation is the Cartan--Schouten--Coxeter given by
\be
f_{abc}=1\quad\text{for}\quad(abc)\in\{(124), (235), (346), (457), (561), (672), (713)\},
\ee
while the remaining ones can be obtained from the aforementioned properties. The~commutator of two octonions is 
\be
[\eo_a,\eo_b]=2f_{abc}\eo_c.
\ee

Unlike its lower dimensional relatives, the~octonions are not associative, and this property is encoded in the so-called associator
\be
[\eo_a,\eo_b,\eo_c]=(\eo_a\eo_b)\eo_c-\eo_a(\eo_b\eo_c)\neq0.
\ee

In the seven-dimensional space of the pure imaginary octonions, we can also introduce the dual of the structure constants
\be
\tilde{f}_{abcd}=\frac{1}{6!}\epsilon_{abcdefg}f_{efg}
\ee
which are related to the non-associative character of $\Oc$. The~analogous dual of the structure constants of $\Ha$ vanish identically by virtue of the Jacobi identity $[\eh_i,[\eh_j,\eh_k]]=0$ that reflects the associative property of the quaternionic algebra. For~$\Oc$, we instead have $[\eo_a,[\eo_b,\eo_c]]=3\tilde{f}_{abcd}\eo_d$ that reveals the untamed character of octonions who do not even comply with~associativity. 

Due to their non-associative nature, octonions have remained even more neglected than their more amicable relatives the quaternions. For~this, the~unit octonions do not even form a group, while unit quaternions are keen to be related to physically sound groups such as rotations or Lorentz transformations. They however hide a beautiful gem inside their more intricate algebraic structure that is unveiled once we \footnote{With a little help from E. Cartan~\cite{Cartan1915}.} note that the automorphism group of the octonions is the exceptional Lie group $G_2$, so the structure constants $f_{abc}$ and their dual $\tilde{f}_{abcde}$ are invariants of $G_2$. In~this sense, the~octonions follow an analogy with the quaternions since the Lie algebra of $G_2$ can be represented in terms of pure imaginary octonions (and $G_2$ can be obtained via exponentiation). Furthermore, one can construct an $SU(3)$ subgroup of $G_2$ as the little group of some imaginary octonionic unit. This is the joyful moment when we realise that octonions may be willing to cooperate for describing color, as~explored for instance by G\"uydin and G\"ursey in~\cite{GuydinandGursey}.

After discovering the octonions and their potential suitability to describe quarks, nobody can blame us for further pursuing our algebra hunt. A~useful result at this point is that $\R$, $\Z$, $\Ha$ and $\Oc$ can be sequentially generated by means of the Cayley--Dickson algorithm. However, the~application of the Cayley--Dickson algorithm to $\Oc$ leads to the so-called sedenions that lack a well-defined division. While it is true that in each iteration we give up one important property (order, commutativity and associativity for $\Z$, $\Ha$ and $\Oc$ respectively), not having a division may seem an excessive concession so one could be a bit more reluctant to include them in the class of {\it{ sensible numbers}
}. \footnote{Of course, this has not prevented to find physical applications for the sedenions as recently explored in e.g.,~\cite{Masi:2021cgm}.}

\begin{figure}[H]
\begin{center}
\includegraphics[width=0.7\linewidth]{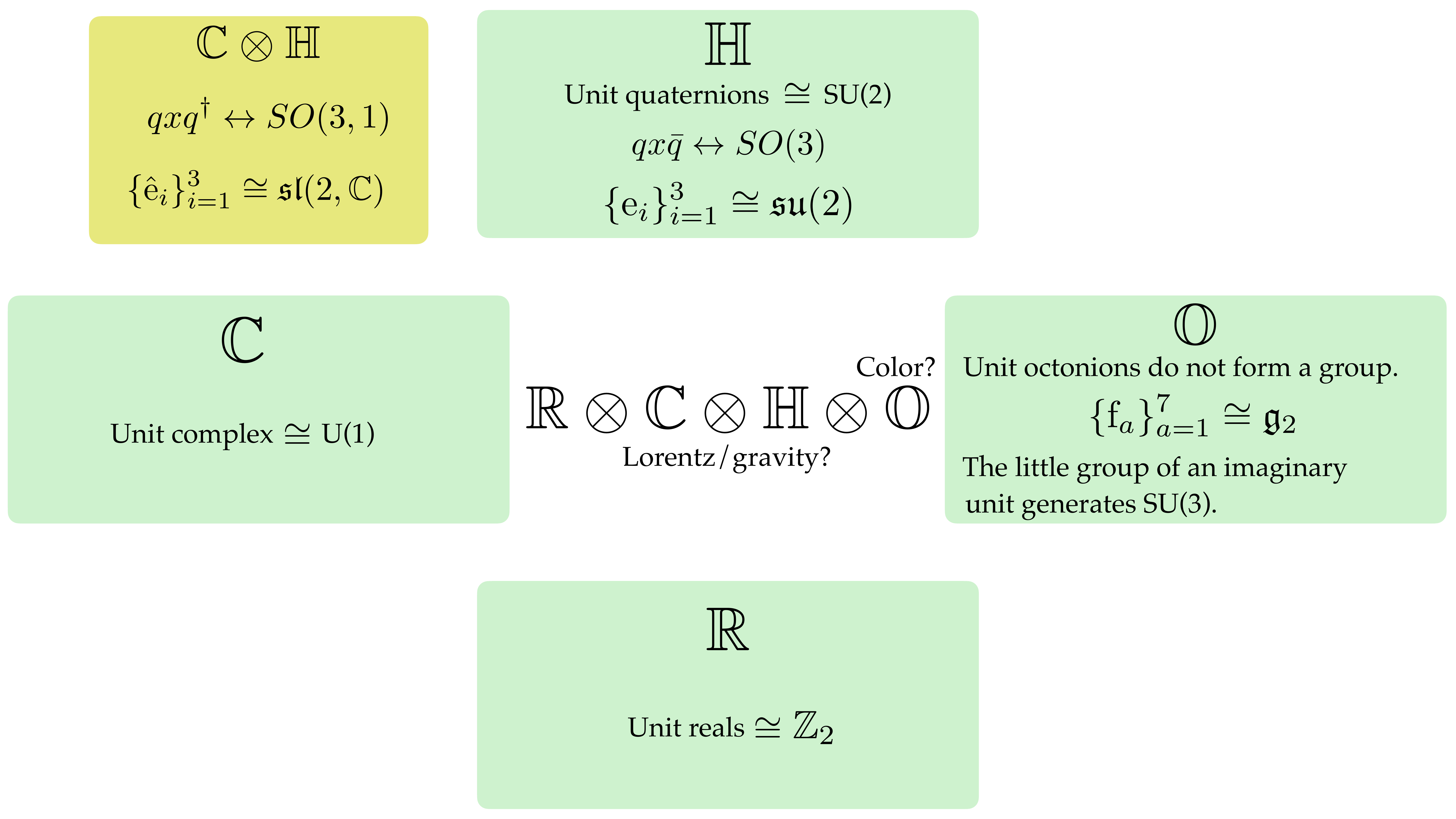}
\end{center}
\caption{In {this} 
 diagram, we summarise the relations of the different algebras with physically relevant groups as explained in the main~text.}
\label{fig:algebras}
\end{figure}

Thus, it seems we have to content ourselves with the reals, the~complex, the~quaternions and the octonions as our possible numbers. This is supported by Hurwitz's theorem that states that the only normed division algebras are indeed $\R$, $\Z$, $\Ha$ and $\Oc$. This is a very remarkable result and, thus, it is very tempting to find their appropriate place to describe physics. Delving into their respective group structures, we have seen that the real unit quaternions are nicely related to $SU(2)$, which is a~result that resembles the relation of unit complex numbers to $U(1)$, while complex quaternions naturally give representations of the Lorentz group. On~the other hand, octonions are related to $G_2$, which contains $SU(3)$ as a subgroup. Thus, we conclude that the only four normed division algebras are intimately related to the fundamental groups of the standard model (possibly even including gravity). It is then extremely appealing to employ \footnote{\textls[-25]{It is clear that the factor $\R$ is redundant and considering $\Z\otimes\Ha\otimes\Oc$ would be sufficient. We prefer to keep it there to emphasise that the general framework of the only four normed division algebras is employed}.} $\algebras$ as the algebraic framework to formulate the fundamental laws of physics (see Figure \ref{fig:algebras}). Exploring these speculations has led to remarkable insights, and it is an increasingly viable hypothesis that the elementary forces and particles are numbers (see e.g.,~\cite{Dixon1994,Furey:2015yxg,Gording:2019srz}).

\section{Algebras in the~Sky}

\textls[-25]{We have discussed how the algebras can provide a very appealing framework for the standard model of elementary particles and, in~particular, the~intriguing suitability of $\algebras$ to describe the particles. In~this essay, we want to explore the potential relevance for cosmology. The~first two factors of the algebra have been extensively applied to cosmological models where real and complex fields have a long history. We will thus focus on the last two factors that remain nearly unexplored (see e.g.,~\cite{Gunaydin:2020ric} for a recent application). The~interest is not only in that they have been seldom studied in cosmological scenarios, but, as~we will argue in the following, they happen to permit more interesting realisations of the cosmological symmetries dictated by the cosmological principle as the usual homogeneity and isotropy \footnote{Cosmological models violating these symmetries have also been considered as in the Bianchi or Lema\^itre-Tolman-Bondi cosmologies.}. This requirement in turn forces our universe to have maximally symmetric spatial sections so they could be described by the groups $ISO(3)$, $SO(3,1)$ and $SO(4)$ for the flat, open and closed universes, respectively. Our universe seems to prefer being flat, so we will assume $ISO(3)$. Thus, in~order to develop cosmological models, we need to have a residual $ISO(3)$ symmetry for our background configuration of fields. The~reals and the complex numbers give trivial realisations of this symmetry group because both real and complex scalar fields comply with these symmetries by simply taking a homogeneous vacuum state. In~that case, homogeneity and isotropy are obvious and they are generated by the usual linear and angular~momentum. }

Moving on to the higher dimensional algebras, things become more interesting. Before~proceeding, it is convenient to pause for a moment and describe the mechanism of the cosmological scenarios based on quaternions and octonions with a more familiar framework. The~underlying idea that we will exploit consists of realising the cosmological principle not directly in terms of the Euclidean subgroup of the Poincar\'e symmetry but~as a diagonal subgroup of $ISO(3,1)\times G$, where $G$ is some internal group. Thus, homogeneity and isotropy are generated by linear and angular momenta that arise as some linear combinations of those within $ISO(3,1)$ and some generators of $G$. Of~course, not every $G$ allows for this construction, but~it should contain some subgroup isomorphic to rotations plus translations. There are many different manners in which this symmetry breaking pattern can be realised, and we refer to~\cite{Nicolis:2015sra} for an exhaustive and comprehensive classification. In~this respect, we only need to resort to the natural symmetry groups that these structures are endowed with. For~the quaternions over the reals, this requirement means that the resulting theory will naturally have an $SU(2)$ symmetry, since this is the symmetry group of unit quaternions as well as the symmetry group of automorphisms of $\Ha$. With~this rudimentary algebra, we can already start considering interesting cosmological applications. Let us proceed to explore some of them:

\begin{itemize}

\item {\it Quaternionic solid inflation}. A~scenario with a scalar quaternionic field $\phi$ can adopt a profile of the form $\langle\phi\rangle=x^i\eh_i$ that breaks isotropy as well as the quaternionic $SU(2)$ symmetry. However, there is a linear combination that is preserved. Regarding homogeneity, we need to make an additional assumption of some internal Abelian symmetry $T$ (e.g., a~shift symmetry) that could restore homogeneity. The~symmetry breaking pattern would then be $ISO(3,1)\times SU(2)\times T\rightarrow ISO_{\text{D}}(3)$. This symmetry breaking pattern appears in solid inflation~\cite{Endlich:2012pz}, and our construction provides a quaternionic formulation of this scenario. {Let us elaborate on the quaternionic formulation of the solid. We can resort to a pure imaginary quaternionic field $\phi(x)$, thus satisfying $\bar{\phi}=-\phi$, so that (global quaternionic) rotations can be realised with unit quaternions. Thus, the~simultaneous action of a spatial rotation $x^i\to R^i{}_jx^j$ and a quaternionic rotation $\phi(x)\to r\,\phi(x)\,\bar{r}$ on $\langle\phi\rangle$ yields 
\be
\langle\phi\rangle=x^i\eh_i\to \langle\tilde{\phi}\rangle=r\left(R^i{}_jx^j\right)\eh_i\bar{r}=x^j\left(\,r\,R^i{}_j\,\eh_i\,\bar{r}\right).
\ee

Now, we can choose $r$ so that the imaginary quaternion $R^i{}_j\,\eh_i\,$ is rotated to $\eh_j$ and, therefore, $\langle\phi\rangle$ remains invariant thanks to the cooperation of $R^i{}_j$ and $r$. In~order to formulate the solid theory, we need to write down a Lagrangian that is required to be real and enjoy Lorentz invariance and both (global quaternionic) rotations and shift symmetry. We will not delve much into the procedure to systematically construct the allowed terms. Instead, we will simply quote that the required conditions are fulfilled by the quaternionic operators: 
\be
X=\partial_\mu\phi\partial^\mu\bar{\phi},\quad Y=\partial_\mu\phi\partial_\nu\bar{\phi} \partial^\mu\phi\partial^\nu\bar{\phi}\quad \text{and}\quad Z=\partial_\mu\phi\partial_\nu \bar{\phi} \partial_\rho\phi\partial^\mu\bar{\phi}\partial^\nu\phi\partial^\rho\bar{\phi}.
\ee

It is straightforward to see that they are real, shift symmetric, Lorentz invariant, and~have the symmetry $\phi\to r\,\phi\,\bar{r}$. It is less trivial to see that they exhaust all the possibilities, so any other operator satisfying the desired properties is a function of the above three. In~fact, $X$, $Y$ and $Z$ encode the three independent invariants of the matrix $\hat{B}$ with components $B^{ij}\equiv \partial_\mu\phi^i\partial^\mu\phi^j$ where $\phi(x)=\phi^i(x)\eh_i$. It can be shown, with~a tedious but simple direct computation, that $X$, $Y$ and $Z$ are linear combinations of the traces of $\hat{B}$, $\hat{B}^2$ and $\hat{B}^3$, which are equivalent to the three fundamental objects employed in~\cite{Endlich:2012pz}.}

\item {\it Triad cosmology}. Let us now consider a real quaternionic \footnote{Let us pedantically clarify that by real quaternionic field we refer to a vector field over $\Ha$ with real coefficients.} vector field $\mathcal{Q}_\mu$ that takes a background configuration of the form $\langle\mathcal{Q}\rangle=Q(t)\delta^a_{i} \eh_a \text{d} x^i$ with $Q(t)$ some time-dependent function.  Again, the~quaternionic sector naturally introduces an $SU(2)$ symmetry that can conspire with the spacetime Lorentz symmetry to preserve a diagonal $SO(3)$. The~scenario is now analogous to models with multiple vector fields featuring internal non-Abelian (global or local) symmetries (see e.g.,~\cite{BeltranJimenez:2018ymu} and references therein).

\item {$\algebras$ \it{cosmology}}. In~the previous examples, it was necessary to resort to some external elements that could assist us in developing the non-trivial realisations of homogeneity and/or isotropy. For~the quaternionic solid, an~additional shift symmetry was necessary, while the quaternionic triad is required to use quaternionic vector fields. It is however possible to overcome these seemingly assisting structures and simply embrace the full generality of $\algebras$. As~explained above, the~complex quaternionic sector conveniently contains Lorentz, while the octonionic sector can account for internal symmetries such as color. Thus, a~theory using this full algebra will have a symmetry group $G$ that will contain, at~least, the~natural groups of the algebra. We have seen that the group of automorphisms of the quaternions and the octonions are $SU(2)$ and $G_2$ respectively. Furthermore, the~complex quaternions provide representations of the Lorentz group. In~this case, we can envision having \linebreak $G\supset SO(3,1)\times G_2$ as the symmetry group, so it is clear that a symmetry breaking pattern $G\rightarrow ISO_{\rm D}(3)$ is possible, where the linear and angular momenta generating this residual three-dimensional Euclidean group is a combination of the generators in $SO(3,1)$ and $G_2$. This group contains several $SU(2)$ subgroups so, in~particular, the~triad cosmology explained above in terms of vector quaternions is possible. However, other realisations are also possible.
\end{itemize}

Let us point out that our reasoning is fully general and the mechanism does not necessitate any specific Lagrangian, but~it is the own structural properties of the algebras that allows the non-trivial realisations of the cosmological principle. These structural properties are enough to obtain observational~signatures.

\section{The sound of cosmic~algebras}

The non-trivial realisations of the cosmological principle provided by quaternions and octonions is not of purely mathematical interest, but~they can have an observational impact for gravitational waves (GWs) astronomy. The~underlying reason is that the background rotational symmetry now combines the usual spatial rotations of space and the internal symmetries of the scalar quaternion and octonion fields. This means that the perturbations will organise themselves into irreps of this symmetry and, as~a consequence, we can have additional helicity-2 perturbations even if the only properly spin-2 field in the theory corresponds to the usual GWs. These additional helicity-2 modes will exhibit a mixing with GWs mediated by the cosmological background configuration of the non-commutative sector of $\algebras$. Depending on the structure of the original group, there could be several additional helicity-2 modes, but~we will focus on the case where only one extra species arises. The~helicity-2 sector will then be conformed by the usual GWs $h_{(\lambda)}$ together with the additional guy $t_{(\lambda)}$ where $\lambda$ stands for the two polarisation modes of each~perturbation.

\begin{figure}[H]
\includegraphics[width=0.95\linewidth]{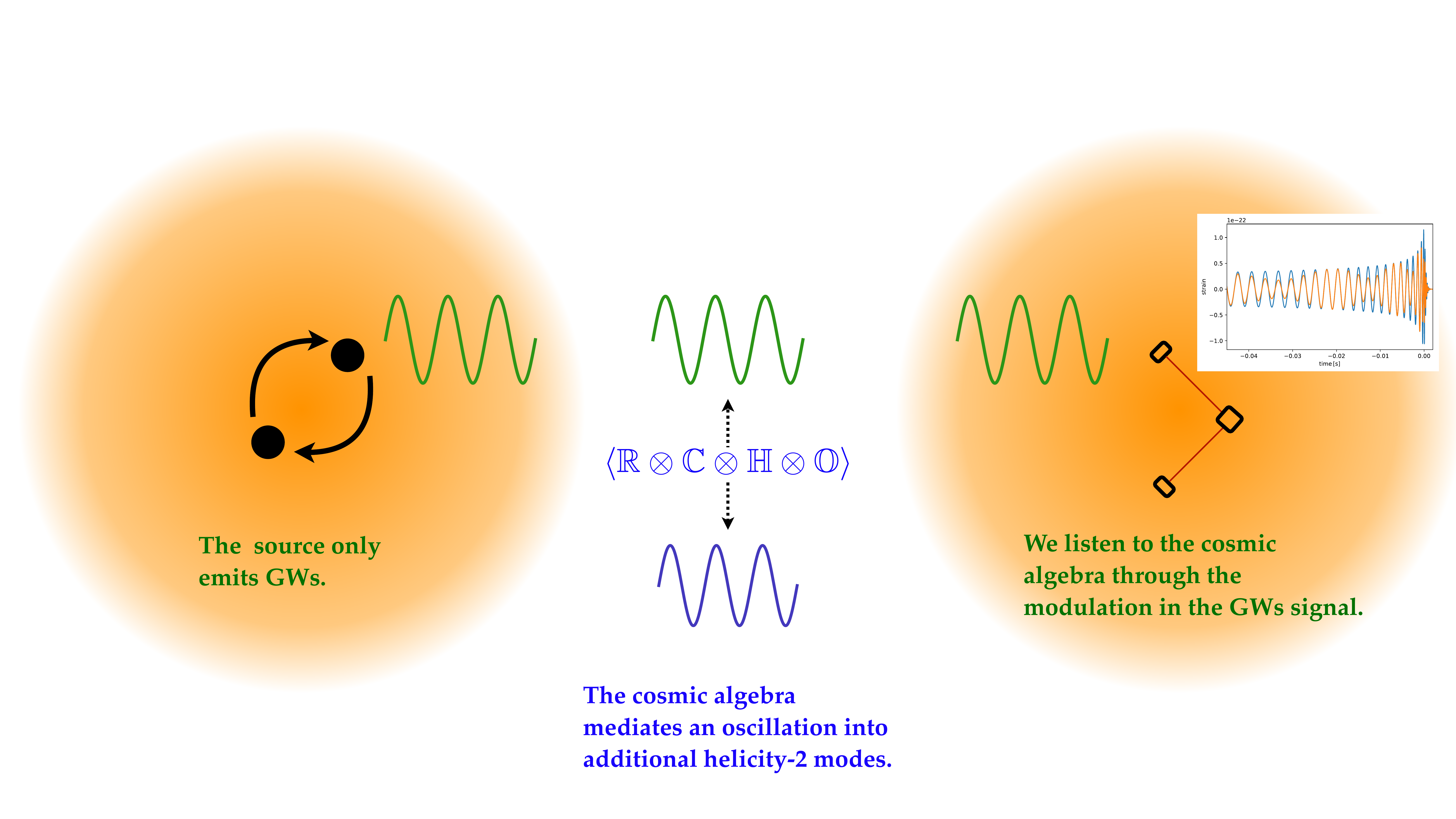}
\caption{This figure illustrates the considered scenario leading to the oscillation of GWs. Initially, a~pure GWs signal is generated by the source. The~cosmic medium with a non-trivial background of $\algebras$ induces an oscillation between GWs. Our interferometers will only be sensitive to the GWs signal, but~the oscillations are imprinted so we can indirectly hear the sounds of the~algebras.}
\label{fig:propagation}
\end{figure}

In the described framework, we can test the presence of a non-trivial background for the non-commutative sector of the algebra $\algebras$ by studying the GWs signal emitted by a binary black hole system. The~scenario we envision is depicted in Figure~\ref{fig:propagation} and consists of two black holes that are assumed to live on an uncharged sector concerning the non-commutative piece of $\algebras$. In~that situation, the~inspiral black holes will only emit the usual GWs $h_{(\lambda)}$, but~no emission in the $t_{(\lambda)}-$channel will be present. As~these GWs travel toward the Earth, they propagate on the non-trivial background of $\algebras$ that will mediate an oscillation into $t_{(\lambda)}$ modes, thus modulating the received signal in our GWs interferometers. We thus arrive at the leitmotif of this essay: {\it {the cosmic presence of $\algebras$ can be heard through GWs}.}

\begin{figure}[H]
\includegraphics[width=0.49\linewidth]{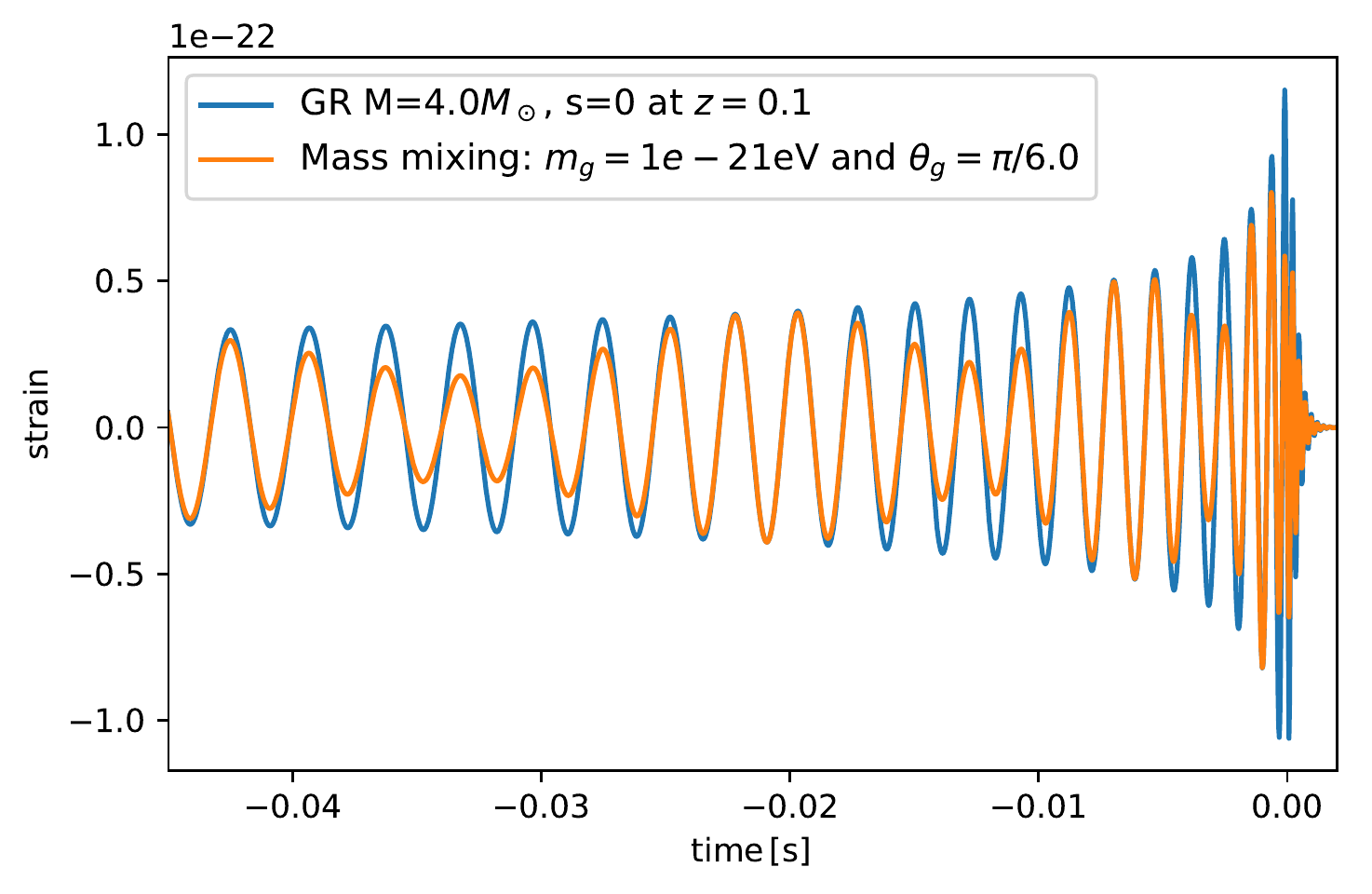}
\includegraphics[width=0.49\linewidth]{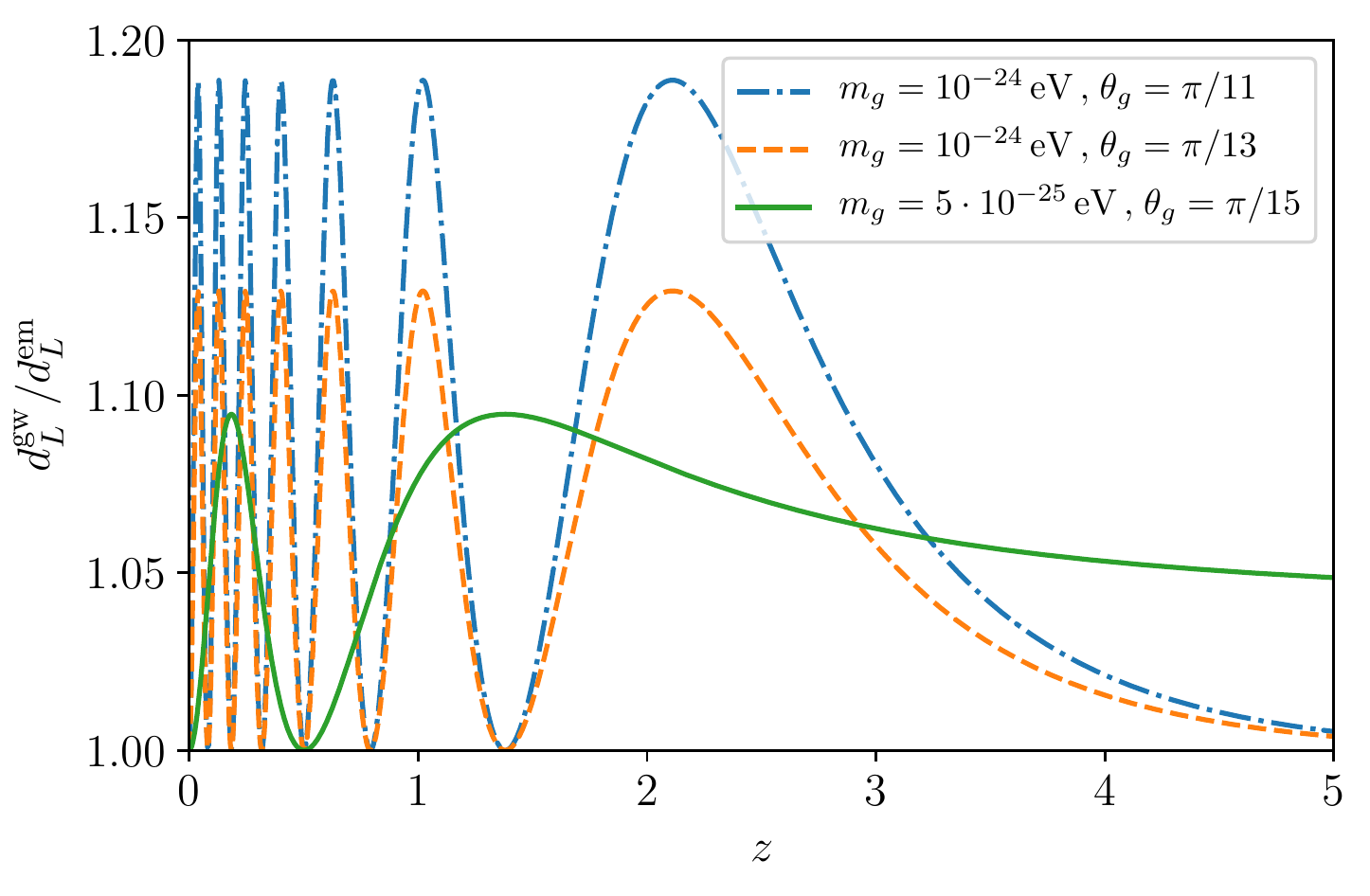}
\caption{{From} 
~\cite{BeltranJimenez:2019xxx}. {The} 
~left panel shows a clear signature of the mixing of GWs with an additional helicity-2 mode mediated by a non-trivial background due to $\algebras$. The~right panel shows a possible signature in the luminosity distance. The~plots correspond to different masses and angle~mixing.}
\label{fig:signatures}
\end{figure}

Under very general assumptions, the~propagation from the source to the receiver will be governed by a system of equations that can be parameterised in Fourier space and in conformal time $\eta$ as
\be \label{eq:generalequation}
\left[\frac{\d^2}{\d\eta^2} + \nuM\frac{\d}{\d\eta}+\cM k^2 + \pM k +\mM\right]\begin{pmatrix} h_{(\lambda)} \\ t_{(\lambda)} \end{pmatrix}=0\,,
\ee
with $k$ representing the Fourier mode and $\nuM$, $\cM$, $\pM$ and $\mM$ representing some matrices encoding the non-trivial cosmological background of $\algebras$. These matrices will typically evolve in time over cosmological time-scales and could also depend on the helicity mode $\lambda$. The~off-diagonal components of these matrices describe the {\it flavor oscillations} that can occur in a variety of manners and lead to different effects, all of which will give rise to distinctive modulations of the GWs signals measured by the interferometers. A~detailed quantitative analysis of the different effects derived from Equation~\eqref{eq:generalequation} can be found in~\cite{BeltranJimenez:2019xxx,Ezquiaga:2021ler}, and some particular cases of GWs oscillations have also been explored for cosmological gauge fields~\cite{Caldwell:2017sto,Caldwell:2018feo} and in massive gravity~\cite{Narikawa:2014fua,Max:2017flc}, though~in this latter case, there is an additional spin-2 field in the theory. Qualitatively, the~flavour oscillations produce interesting effects, some of which we mention in the following (see also Figure~\ref{fig:signatures}):
\begin{itemize}
    \item Anomalous propagation speed of GWs. If~$\cM$ is different from the identity, then the oscillations will induce an anomalous propagation speed of GWs even if $\cM_{hh}=1$. The~reason is that the oscillation will make the GWs propagate as $t$-modes for some time, thus modifying the effective propagation speed throughout its path.
    \item Generation of chirality. If~$\pM$ is different from 0, then the different helicities will oscillate in a different manner, thus generating chirality for the GWs. The~reason is that $\pM$, governing a term linear in $k$, typically arises from violations of parity.
    \item Oscillations in the GWs luminosity distance. Due to the oscillations and the presence of the friction matrix $\nuM$, the~luminosity distance of GWs $d^{\text{GW}}_{L}$ will be affected. Comparing this quantity with the electromagnetic counterpart $d^{\text{EM}}_{L}$, we can also hear some non-trivial cosmic $\algebras$.
\end{itemize}

There are other interesting observational effects that will be visible in GWs astronomy such as echoes, wave distortions or birrefringence that we will not explain in detail but~are comprehensively analysed in~\cite{BeltranJimenez:2019xxx,Ezquiaga:2021ler}. The~remarkable result is that the subtle whisper of the cosmic algebras will be imprinted in the GWs, opening the possibility to test their presence through a variety of~effects.

\unskip

\section{Conclusions}

Normed division algebras are interesting fellows from a pure mathematical viewpoint since they conform {\it {sensible}} types of numbers. Numbers are allegedly our way of communicating with Nature, and the reals and complex numbers have repeatedly proved their appropriateness to that purpose and so they are hardwired in our description of the physical laws. Quaternions have also claimed their position in this endeavour, but~they have remained largely marginalised. They are starting to receive well-deserved attention due to their suitability to describe some fundamental aspects of particles. Octonions are by far the most obscure member of this family, although~they hide extremely remarkable properties that not only are suitable for color but~also permit giving a fresh new look at known intriguing results in, e.g.,~string theory and supersymmetry. This essay has been devoted to reviewing this family and some of its applications in physics. If~this family does play a fundamental role in the foundations of our standard model, including gravity, it could also plausibly participate in the cosmological evolution of our universe. In~this respect, we have argued how the two non-commutative members naturally lead to interesting realisations of the cosmological principle that arise from their structural properties. If~that is the case, they can be probed with different distinctive signatures in GWs astronomy so the future observations of GWs could unveil the presence of cosmic quaternions and octonions and bring exciting news about these undeservedly neglected~acquaints.

We will conclude with a wild surmise. Fundamental aspects of the standard model are encoded in scattering amplitudes and, in~particular, the~analytical properties of physical observables such as the $S-$matrix. In~this respect, complex analysis emerges as a fundamental tool, where analiticity, poles, branch cuts, etc., have precise physical meanings. We cannot resist speculating that quaternionic and octonionic analysis (or analysis in $\algebras$ more generally) could eventually bring out new insights on our comprehension of the most fundamental laws of physics and could resolve long-standing difficulties with the gravity~sector.

\vspace{6pt}


\funding{J.B.J. was supported by the {\it Atracci\'on del Talento Cient\'ifico en Salamanca} programme and the Project PGC2018-096038-B-100 funded by the Spanish ``Ministerio de Ciencia e Innovación''. T.S.K. was supported by the Estonian Research Council grants PRG356 ``Gauge Gravity'', MOBTT86 and by the EU through the European Regional Development Fund CoE program TK133 ``The Dark Side of the Universe''.}

\end{paracol}
\newpage
\reftitle{References}

\bibliography{Refs}

\end{document}